\documentclass[preprintnumbers,amsmath,amssymb,floatfix,prd,12pt,superscriptaddress,nofootinbib]{revtex4}

\def\be {\begin{equation}}
\def\ee {\end{equation}}
\def\ba {\begin{eqnarray}}
\def\ea {\end{eqnarray}}

\def\L  {\Lambda}

\def\bi {\begin{itemize}}
\def\ei {\end{itemize}}
\begin{document}

\title{\textbf{Holographic dark energy in Brans-Dicke cosmology with chameleon scalar field
}}

\author{M. R. Setare} \email{rezakord@ipm.ir}
\affiliation{Department of Science of Bijar, University of
Kurdistan, Bijar, Iran}

\author{ Mubasher Jamil} \email{mjamil@camp.nust.edu.pk}
\affiliation{ Center for Advanced Mathematics and Physics, National
University of Sciences and Technology, H-12, Islamabad, Pakistan}

\begin{abstract}
\vspace*{1.5cm} \centerline{\bf Abstract} \vspace*{1cm} We study a
cosmological implication of holographic dark energy in the
Brans-Dicke gravity. We employ the holographic model of dark energy
to obtain the equation of state for the holographic energy density
in non-flat (closed) universe enclosed by the event horizon measured
from the sphere of horizon named $L$. Our analysis shows that one
can obtain the phantom crossing scenario if the model parameter
$\alpha$ (of order unity) is tuned accordingly. Moreover, this
behavior is achieved by treating the Brans-Dicke scalar field as a
Chameleon scalar field and taking a non-minimal coupling of the
scalar field with matter. Hence one can generate phantom-like
equation of state from a holographic dark energy model in non-flat
universe in the Brans-Dicke cosmology framework.
\end{abstract}

 \maketitle
\newpage
\section{Introduction}

Many cosmological observations, such as SNe Ia \cite{1}, WMAP
\cite{2}, SDSS \cite{3}, Chandra X-ray observatory \cite{4}, etc.,
reveal that our universe is undergoing an accelerating expansion. To
explain this cosmic positive acceleration, mysterious dark energy
has been proposed. There are several dark energy models which can be
distinguished by, for instance, their equation of state (EoS)
$(w_\L=\frac{P_\L}{\rho_\L}) $ during the evolution of the universe.
Astrophysical data also indicate that $w_\L$ lies in a very narrow
strip close to $-1$. The case $w_\L=-1$ corresponds to the
cosmological constant. For $w_\L$ less than $-1$ the phantom dark
energy \cite{10} is observed, and for $w_\L$ more than $-1$ (but
less than $\frac{-1}{3}$) the dark energy is described by
quintessence \cite{quintessence}.  More ever, the analysis of the
properties of dark energy from recent observational data mildly
favor models of dark energy with $\omega$ crossing $-1$ line in the
near past. So, the phantom phase equation of state with $\omega_\L <
-1$ is still mildly allowed by observations. Most of dark energy
models treat scalar field(s) as dark component(s) with a dynamical
equation of state. So far, a large class of scalar-field dark energy
models have been studied, including quintessence
\cite{quintessence}, K-essence \cite{kessence}, tachyon
\cite{tachyon}, phantom \cite{10}, ghost condensate
\cite{ghost1,ghost2} and quintom \cite{quintom}, and so forth. The
introduction of a scalar field potential augmented by a scalar field
dependent coupling constant solved many problems and provided clues
to the solutions of many of the outstanding cosmological problems in
particular within the framework of the Brans-Dicke (BD) cosmology.

In recent years, the holographic dark energy has been studied as a
possible candidate for dark energy. It is motivated from the
holographic principle which might lead to the quantum gravity to
explain the events involving high energy scale. In the
thermodynamics of black hole, there is a maximum entropy in a box of
length $L$, commonly termed, the Bekenstein-Hawking entropy bound
$S\sim M_p^2L^2$ ($c=\hbar=1$), which scales as the area of the box
$A\sim L^2$ \cite{bek}. To avoid the breakdown of the local quantum
field theory, Cohen et al \cite{cohen} suggested that the entropy
for an effective field theory should satisfy $L^3\Lambda^3\leq
S^{3/4}\sim (M_p L)^{3/2}$. Here $L$ is the size of the region which
serves as the infra-red cut-off while $\Lambda$ is the ultra-violet
cut-off. Incidently this last equation can be re-written in the form
$L^3\rho_\Lambda\leq L M_p^2$, where $\rho_\Lambda\sim\Lambda^4$ is
the energy density corresponding to the zero-point energy and
cut-off $\Lambda$. Thus the total energy in a region of size $L$
cannot exceed the mass of a black hole of the same size. From this
discussion we can deduce that $\rho_\Lambda\leq M_p^2L^{-2}$. This
inequality can be saturated and it becomes $\rho_\Lambda=3c^2
M_p^2L^{-2}$, where $3c^2$ is introduced for convenience.

It is well-known that general relativity although performs fairly
well at the solar system scale but fails at the cosmic scales at
which it is unable to explain the origin of dark energy or present
accelerated cosmic expansion. This has lead scientists to propose
alternative theories of gravity. Some of these gravity theories are
actually modifications of Einstein's general relativity including
scalar-tensor gravity \cite{scalar} and $f(R)$ gravity
\cite{gravity}. The earliest modifications to Einstein's gravity was
the Kaluza-Klein gravity \cite{kaluza} which was intended to unify
gravity with the electromagnetic force. In the early 1960's, Brans
and Dicke proposed a relativistic theory of gravity based on the
Einstein's theory of general relativity \cite{dicke}. This was the
first gravity theory in which the dynamics of gravity were described
by a scalar field while spacetime dynamics were represented by the
metric tensor. Later on it turned out that the theory can pass the
experiments from the solar system \cite{bert}. An attractive feature
of BD theory is that the scalar field is a fundamental element of
the theory, quite contrary to other models in which scalar field is
introduced separately in an ad hoc manner. The basic notion is that
the BD scalar field plays the role of quintessence or K-essence and
lead to cosmological acceleration. However it happens only in
certain cases: for instance the cosmic acceleration is permissible
if the BD parameter is constrained in the range $-2<w<-3/2$, which
not only violate the energy conditions on the scalar field but are
also inconsistent with a radiation dominated epoch, unless $w$ is a
time dependent parameter.

As discussed earlier that the scalar field plays the role of dark
energy in the BD gravity. In this connection, we can relate the
energy density of holographic dark energy (in fact, any other form
of dark energy in general) and the BD scalar field in a consistent
way. The holographic dark energy in the framework of Brans-Dicke
gravity has been investigated in \cite{set,feng}. We consider a
Brans-Dicke framework in which there is a non-minimal coupling
between the scalar field and the background geometry. The action is
modified due to the inclusion of a non-minimal coupling of the
scalar field with the matter. This work is actually motivated by the
work of Clifton and Barrow \cite{barrow} where they investigated the
behavior of an isotropic cosmological model in the early as well as
in the late time limits in the Brans-Dicke framework. However, the
non-minimal coupling of a scalar field with both of geometry and the
matter sector has been in use for quite a long time, for instance,
in the dilaton gravity \cite{dilaton}.

\section{Holographic Energy Density in Brans-Dicke Framework}

We start with the action for the Brans-Dicke theory \cite{das}
\begin{equation}
A=\int\sqrt{-g}d^4x\Big[-\Phi
R+\frac{\omega}{\Phi}\Phi^{,\mu}\Phi_{,\mu}+f(\Phi)L_m \Big],
\end{equation}
in the Jordan frame. Here $R$ is the Ricci scalar and $\Phi=\Phi(t)$
is the Brans-Dicke scalar field representing the inverse of Newton's
constant which is allowed to vary with space and time and $\omega$
is the generic dimensionless parameter of the Brans-Dicke theory.
The matter Lagrangian $L_m$ represents the perfect fluid matter.
Notice that taking $\omega=0$ and $f=0$ in (1) yields the
Einstein-Hilbert action. In the Jordan frame, the matter minimally
couples to the metric and there is no interaction between the scalar
field $\Phi$ and the matter field \cite{gong}. Here $\omega$ is an
unknown parameter.

We assume the background to be Friedmann-Robertson-Walker (FRW)
spacetime
\begin{equation}
ds^2=-dt^2+a(t)^2\Big[\frac{dr^2}{1-kr^2}+r^2(d\theta^2+\sin^2\theta
d\phi^2) \Big].
\end{equation}
Here $a$ is the scale factor while $k=-1,0,+1$ corresponds to open,
flat and closed universe. The gravitational field equations derived
from the variation of the action (1) with respect to FRW metric is
\begin{eqnarray}
H^2+\frac{k}{a^2}&=&\frac{f}{3\Phi}\rho+\frac{1}{6}\omega
\frac{\dot{\Phi}^2}{\Phi^2}-H\frac{\dot\Phi}{\Phi},\\
2\frac{\ddot{a}}{a}+H^2+\frac{k}{a^2}&=&p-\frac{\omega}{2}
\frac{\dot{\Phi}^2}{\Phi^2}-\frac{\ddot{\Phi}}{\Phi}
-2H\frac{\dot\Phi}{\Phi}.
\end{eqnarray}
Here total energy density $\rho=\rho_\Lambda+\rho_m$, the sum of
energy densities of holographic dark energy and matter. Also the
total pressure $p=p_\Lambda+p_m$, is the sum of pressure of
holographic dark energy and matter. The dynamical equation for the
scalar field is
\begin{equation}
6\Big(\frac{\ddot{a}}{a}+H^2+\frac{k}{a^2}\Big)
+\omega\frac{\dot{\Phi}^2}{\Phi^2}+L_mf_{,\Phi}-
2\omega\frac{\ddot{\Phi}}{\Phi}=0.
\end{equation}
The parameter $\omega$ that appears in the above field equations (3)
to (5) can be observed astronomically. The Cassini experiment
implies that $\omega>10^{4}$ \cite{cassini}. We define parameters as
\begin{equation}
\Omega_k=\frac{2k}{a^2H^2},\ \ \rho_{cr}=6\Phi H^2.
\end{equation}
The holographic dark energy is related to the Brans-Dicke scalar
field by
\begin{equation}
\rho_\Lambda=6\Phi L^{-2},
\end{equation}
where the holographic parameter is fixed at $c=1$. Since there are
numerous candidates of dark energy, it is important to know how
these candidates are related to each other. Notice that from Eq.
(7), the dynamics of BD scalar field are connected with that of the
energy density of HDE. Also note that if the infrared cut-off is
taken as the Hubble horizon then the energy density of HDE and
critical density match identically. This situation generically
arises in inflation scenario where $L=H^{-1}$ (a constant length
scale). However in the overall evolution of the Universe with
$0<c<1$ and $L\neq H^{-1}$, the two energy densities are not the
same and $\rho_\Lambda<\rho_{cr}$.

There are various choices of infrared cut-off for the cosmological
length scale available in the literature. For instance, the simplest
choice for infrared cut-off is the Hubble horizon $H^{-1}$, but Hsu
\cite{hsu} has shown that in this case the EoS parameter
$\omega_\Lambda=0$, which can not describe the cosmic acceleration.
Another suggestion is the particle horizon
($L_p=a(t)\int\limits_0^t\frac{dt'}{a(t')}$), but it turns out that
$\omega_\Lambda>-1/3$, a form of exotic matter that is not suitable
to derive acceleration. A suitable choice of infrared cut-off was
then suggested by Li \cite{li1} and is termed as the future event
horizon (defined below). Note that the choice of horizon (or the
infrared cut-off) is independent of the spatial curvature of FRW
spacetime. According to the formulation of holographic dark energy
in non-flat geometry, the cosmological length $L$ is considered to
be:
\begin{equation}\label{Lnonflat}
L\equiv\frac{a(t)}{\sqrt{|k|}}\,\text{sinn}(\sqrt{|k|}y),\ \ \
y=\frac{R_h}{a(t)}~,
\end{equation}
where $R_h$ is the future event horizon defined by
\begin{equation}
 R_ h(a)=a\int_t^\infty{dt'\over
a(t')}=a\int_a^\infty{da'\over Ha'^2}~,
\end{equation}
and
\begin{equation}\frac{1}{\sqrt{|k|}}\text{sinn}(\sqrt{|k|}y)=
\begin{cases} \sin y  & \, \,k=+1,\\
             y & \, \,  k=0,\\
             \sinh y & \, \,k=-1.\\
\end{cases}\end{equation}
In some recent studies, some new infrared cut-offs have been
proposed. In \cite{23}, the authors have added the square of the
Hubble parameter and its time derivative within the definition of
holographic dark energy. While in \cite{24}, the authors propose a
linear combination of particle horizon and the future event horizon.
However, in this paper we will adopt Li's proposal.

The holographic dark energy satisfies
\begin{equation}
\Omega_\Lambda=\frac{1}{H^2L^2}.
\end{equation}
Differentiating (8), we obtain
\begin{equation}
\dot L=\frac{1}{\sqrt{\Omega_\Lambda}}-\text{cosn}(\sqrt{|k|}y),
\end{equation}
where
\begin{equation}\text{cosn}(\sqrt{|k|}y)=
\begin{cases} \cos y  & \, \,k=+1,\\
             1 & \, \,  k=0,\\
             \cosh y & \, \,k=-1.\\
\end{cases}\end{equation}
The energy conservation equation is
\begin{equation}
\dot{\rho}_\Lambda+3H(\rho_\Lambda+p_\Lambda)=0,
\end{equation}
where $p_\Lambda$ is the pressure of the HDE. Notice that the above
expression (14) for HDE is valid \cite{set,kim} while the continuity
equation for matter is given in \cite{das}. Differentiating (7)
w.r.t. time, we get
\begin{equation}
\dot{\rho}_\Lambda=\rho_\Lambda\Big(
\frac{\dot\Phi}{\Phi}-2\frac{\dot L}{L}  \Big).
\end{equation}
Taking an ansatz for BD scalar field \cite{das}
\begin{equation}
\frac{\dot\Phi}{\Phi}=-\frac{\alpha}{H} ,
\end{equation}
where $\alpha$ is a constant of order unity (and not arbitrary).
There is no a priori physical motivation for this choice, this is
purely phenomenological which leads to the desired behavior of the
deceleration parameter $q$ of attaining a negative value at the
present epoch from a positive value during a recent past. Using (12)
and (16) in (15), we obtain
\begin{equation}
\dot{\rho}_\Lambda=-\rho_\Lambda H\left[ \frac{\alpha} {H^2} +
2\left(1 -\sqrt{\Omega_\Lambda} \text{cosn}(\sqrt{|k|}y)\right)
\right].
\end{equation}
Putting (17) in (14), we get
\begin{equation}
p_\Lambda=-\frac{1}{3}\Big[1-\frac{\alpha}
{H^2}+2\sqrt{\Omega_\Lambda}\text{cosn}(\sqrt{|k|}y)\Big]\rho_\Lambda.
\end{equation}
Thus equation of state parameter of dark energy in Brans-Dicke
cosmology becomes
\begin{equation}
\omega_\Lambda=-\frac{1}{3}\Big[1-\frac{\alpha}
{H^2}+2\sqrt{\Omega_\Lambda}\text{cosn}(\sqrt{|k|}y)\Big].
\end{equation}
Notice that for particular choices of $\alpha$, it is possible to
attain $\omega_\Lambda>-1$, $\omega_\Lambda<-1$ and values in
between. Thus a phantom crossing scenario can be constructed in the
present model.

Now we determine the evolution of $\Omega_\Lambda$. We differentiate
it w.r.t $t$ to get
\begin{equation}
\frac{d \Omega_\Lambda}{dt}=-2\Omega_\Lambda\Big(\frac{\dot
H}{H}+\frac{\dot L}{L}\Big).
\end{equation}
Using the definitions
\begin{equation} \dot H=-H^2(1+q),\ \ \
q=-\frac{\ddot a}{aH^2},
\end{equation}
in Eq. (20), we obtain
\begin{equation}
\frac{d\Omega_\Lambda}{dt}=2\Omega_\Lambda
H\Big[q+\sqrt{\Omega_\Lambda}\text{cosn}(\sqrt{|k|}y)\Big].
\end{equation}
Here $q$ is the deceleration parameter. From Eq. (4), it is easy to
check
\begin{equation}
q=\frac{1}{\alpha+2H^2}\Big[H^2\Big(1+\frac{\Omega_k}{2}\Big)+\frac{\alpha^2}{H^2}
\Big(1+\frac{\omega}{2}\Big)-3\alpha-p\Big].
\end{equation}
Notice that if we take $k=0$ and $p=0$, then (23) reduces to the
case discussed by Das and Banerjee \cite{das}, thus our paper
extends their work.

In a spatially flat universe, we have
\begin{equation}
\omega_\Lambda=-\frac{1}{3}\Big(1-\frac{\alpha}
{H^2}+2\sqrt{\Omega_\Lambda}\Big).
\end{equation}
Using $\Omega_\Lambda=0.73$ in (24), we notice that
$\omega_\Lambda>-1$ when $\alpha>-0.2912H^2$, while
$\omega_\Lambda<-1$ when $\alpha<-0.2912H^2$.

To check how the HDE state parameter $\omega_\Lambda$ evolves in a
spatially flat Universe, we calculate its time derivative and get
\begin{equation}
\dot{\omega}_\Lambda(t)=\frac{2}{3H}(1+q)-
\frac{2}{3}H\sqrt{\Omega_\Lambda}(q+\sqrt{\Omega_\Lambda}).
\end{equation}
Notice that in an accelerating Universe, $q\leq-1$, therefore (25)
implies $\dot{\omega}_\Lambda<0$. Hence in HDE dominated flat
Universe, the state parameter evolves gradually to super-negative
values.

In the end, we would comment that the present study can also be
performed for the new agegraphic dark energy (NADE). In NADE, the
infrared cut-off is the conformal time,
$\eta=\int\limits_0^a\frac{da'}{Ha'^2}$, i.e. the age of the
Universe is used as a parameter in the model \cite{ahmad}. In this
context, it will be interesting to see whether the phantom crossing
will take place in both flat and non-flat FRW backgrounds in the
context of BD cosmology. However there are several disadvantages
with the agegraphic dark energy: it can not generate the inflation
era in the early universe unlike the holographic dark energy; it
cannot produce a phantom dominated universe since its equation of
state parameter is always greater than $-1$; its energy density
decreases with time unlike any other dark energy candidate; quantum
corrections are generally ignorable and it worse fits with the
observational data \cite{li,wei1}. From the same study \cite{li}, it
is deduced that the observational data is well-fitted with the
dynamical dark energy models with only one free parameter including
the holographic dark energy model. However the NADE model is
completely disfavored by the astrophysical data. Thus our present
study is also consistent with the empirical findings.

\section{Conclusion}

In this paper, we have studied holographic dark energy in the
Brans-Dicke cosmological model. The later model contains a dynamical
time dependent scalar field that behaves like dark energy and
produces cosmic acceleration. Following the work of \cite{das}, we
assumed the scalar field to be chameleon i.e. a scalar field
interacting with the matter. The two entities are connected to each
other via a coupling parameter $f$. Although we took $f$ to be
arbitrary function, Das and Banerjee \cite{das} took few ansatz for
it and studied the behavior of the model. Ideally, the coupling
between dark energy (of any type including the BD scalar field) and
matter should be derived from a theory of quantum gravity. In the
absence of such a theory, we decided to keep our analysis general
regardless of the specification of $f$. The coupling parameter could
also be constrained from the astrophysical data.

Earlier it had been observed in \cite{set} that if the BD scalar
field is minimally coupled to gravity then phantom crossing is not
possible. In that model, the author took the BD scalar field as
related with the HDE. The study of \cite{set} motivated us to
consider a modification of the BD framework by introducing a
chameleon scalar field interacting with the matter. By establishing
a correspondence between the BD scalar field and the HDE, we have
found that phantom crossing is possible by tuning the free parameter
$\alpha$. Notice that the parameter $\alpha$ is not arbitrary but is
a constant of order unity. We would also comment that the fine
tuning of parameters is a general feature of almost every
cosmological model and our model is also no exception. Finally, we
took the infrared cut-off of the cosmological spacetime to be future
event horizon $R_h$ and performed the analysis in both flat and
non-flat backgrounds.

\subsubsection*{Acknowledgment}
Both of us would like to thank the referee for his critical comments
to improve this work.

\end{document}